\documentclass[
    ,final            
  ]
  {aipproc}

\layoutstyle{8x11single}

\usepackage{amsmath,amssymb}
\usepackage{bm}

\begin{document}

\title{A Novel View of Exotic Hadrons and the Covariant
$\bm{\widetilde{U}(12)}$-Classification Scheme}


\author{Kenji Yamada}{
  address={Department of Engineering Science,
 Junior College Funabashi Campus, Nihon University, \\
 Funabashi, Chiba 274-8501, Japan}
}



\begin{abstract}
I classify exotic hadrons into two types, ``genuine'' and ``hidden''
 exotics, and propose that the ``hidden'' exotics would be
 interpreted as ``chiralons'' in the $\widetilde{U}(12)_{SF}
 \times O(3,1)_{L}$-classification scheme of hadrons. Based upon this
 conjecture I investigate the mass spectrum of $1S$, $1P$, and $2S$ states
 for the $c\bar{c}$ system by use of a phenomenological mass formula
 with the spin-dependent interactions and present
 possible assignments for exotic neutral charmonium-like states
 which were recently discovered. It is also mentioned that some
 $J^{PC}$-exotic states predicted in this scheme might correspond
 to those found in a recent lattice QCD calculation of the charmonium
 spectrum.
 
\end{abstract}

\maketitle


\section{Introduction}

In recent years there have been discovered numerous exotic states
 which have unexpected and puzzling nature, such as charmonium-like
 states, $XYZ$ series, and $\theta(1540)$, and also there seem to be
 too many observed mesons, especially in the light $(u,d)$-quark sector, to
 be classified as the conventional
 quark-model $q\bar{q}$ states \cite{PDG2009}. These experimental
 situations have revived multiquark hadrons and their theoretical
 studies which were declining before them. A serious complication
 of most models suggesting multiquark states is that they would
 predict too many states to exist, that is, the known problem of
 ``state inflation'', and further, though many theoretical
 proposals for interpreting exotic charmonium-like states have
 put forward, no single framework of models can accommodate most
 of those states. 


\section{The $\bm{\widetilde{U}(12)_{SF}
 \times O(3,1)_{L}}$-Classification Scheme of Hadrons}

The $\widetilde{U}(12)_{SF} \times O(3,1)_{L}$-classification
 scheme \cite{IIM2000} is a covariant system of classification
 for hadrons which
 gives covariant quark representations for composite hadrons
 with definite Lorentz and chiral transformation properties.
 In the rest frame of hadrons the $SU(2)_{\sigma}$ intrinsic
 spin symmetry is extended to $U(4)_{S}$ by adding a new
 degree of freedom on the $\rho$-spin, $SU(2)_{\rho}$,
 being connected with decomposition of Dirac matrices,
 $\gamma = \rho \times \sigma$. Then an extended $U(12)_{SF}$
 spin-flavor symmetry, including the flavor $SU(3)_{F}$, has
 as its subgroups both the nonrelativistic spin-flavor and
 chiral symmetry, as $SU(6)_{SF} \times SU(2)_{\rho}$ and
 $U(3)_{L} \times U(3)_{R} \times SU(2)_{\sigma}$.
 The $\widetilde{U}(12)_{SF} \times O(3,1)_{L}$-classification
 scheme is obtained through the covariant generalization from
 $U(12)_{SF} \times O(3)_{L}$ by boosts,
 separating the spin and space degrees of freedom.
 Thus the $U(12)_{SF}$ symmetry in the hadron rest
 frame is embedded in the covariant $\widetilde{U}(12)$-representation
 space, which includes subgroups as $\widetilde{U}(4)_{D}
 \times SU(3)_{F}$, $\widetilde{U}(4)_{D}$ being the pseudounitary
 homogeneous Lorentz group for Dirac spinors.

An essential ingredient of the $\widetilde{U}(12)_{SF}
 \times O(3,1)_{L}$-classification
 scheme is that quarks have the $\rho$-spin degree of freedom,
 which is discriminated by the eigenvalues $r=\pm$
 of $\rho_{3}$ in the hadron rest frame. This implies that not only
 conventional quarks with $r=+$ but also new type quarks with
 $r=-$ are building blocks of hadrons. Thus hadron states are
 characterized, aside from flavors, by the quantum numbers,
 the net constituent $\rho$-spin $S^{(\rho)}$, its third component
 $S^{(\rho)}_{3}$, the ordinary net constituent $\sigma$-spin $S$,
 the net orbital angular momentum $L$, and the total spin $J$.
 
For $q\bar{q}$ meson states the meson spin $J$, parity $P$, and
 charge-conjugation parity $C$ are given by
\begin{equation}
	\mathbf{J}=\mathbf{L}+\mathbf{S}, \ \
	P=(-1)^{L+|S^{(\rho)}_{3}|}, \ \
	C=(-1)^{L+S+S^{(\rho)}+1}.
\end{equation}
 The ground ($L=0$) states of $q\bar{q}$ mesons are composed of
 states with the $J^{PC}$ quantum numbers, two pseudoscalars
 $P(0^{-+})$, $P^{(\chi)}(0^{-+})$, two scalars
 $S^{(\chi)}_{A}(0^{++})$, $S^{(\chi)}_{B}(0^{+-})$, two vectors
 $V(1^{--})$, $V^{(\chi)}(1^{--})$, and two axial-vectors
 $A^{(\chi)}(1^{++})$, $B^{(\chi)}(1^{+-})$, where the symbol
 $\chi$ represents at least one of the
 constituents having the negative eigenvalue $r=-$ of $\rho_{3}$,
 referred to as ``chiralons''. There are also orbital and radial excitations
 for each of these eight types of meson. The $\rho$-spin quantum numbers
 of ground states and their excitations are given by
%
%
\begin{equation}
(S^{(\rho)},S^{(\rho)}_{3}) = \begin{cases}
	(1,1)  &\text{for $P$ and $V$ sectors} \\
	(1,-1) &\text{for $P^{(\chi)}$ and $V^{(\chi)}$ sectors} \\
	(1,0)  &\text{for $S^{(\chi)}_{A}$ and $B^{(\chi)}$ sectors} \\
	(0,0)  &\text{for $S^{(\chi)}_{B}$ and $A^{(\chi)}$ sectors}
\end{cases}
	,
\end{equation}
where the $(1,1)$ sector corresponds to the conventional
 $q\bar{q}$ states. Here I simply assume
 that all ground and excited states are physically realized,
 though it is a highly dynamical problem.


\section{A Novel Classification of Exotic Hadrons and Exotic
 Charmonium-like States}

Exotic hadrons are defined, in the conventional quark
 model, as mesonic and baryonic states with quantum numbers
 forbidden for $q\bar{q}$ and $qqq$ states or anomalous 
 features which cannot be explained as those states. Here I
 classify exotic hadrons into the following two types:
\begin{description}
  \item[``Genuine'' exotics]
  mesonic and baryonic states with flavor quantum numbers forbidden
  for the conventional quark-model $q\bar{q}$ and $qqq$ states.
  \item[``Hidden'' exotics]
  mesonic and baryonic states which are constructed by adding extra
  flavor-singlet light-$q\bar{q}$ or gluon components to the 
  conventional $q\bar{q}$ and $qqq$ states, and pure gluonic ones. 
\end{description}
Then I conjecture that the hidden exotics would be interpreted
 as chiralons in the $\widetilde{U}(12)_{SF} \times
 O(3,1)_{L}$-classification scheme. Mesonic $q\bar{q}$ chiralons
 are considered to be asymptotically four quark states
 ($q\bar{q}$ $+$ flavor-singlet light $q\bar{q}$),
 $q\bar{q}$-gluon states ($q\bar{q}$ $+$ gluon(s)), or pure
 gluonic states ($gg$, $ggg$), while baryonic $qqq$ chiralons
 are considered to be asymptotically five quark states
 ($qqq$ $+$ flavor-singlet light $q\bar{q}$) or $qqq$-gluon
 states ($qqq$ $+$ gluon(s)). 

\subsection{The charmonium spectrum in the
 $\widetilde{U}(12)_{SF} \times O(3,1)_{L}$-classification scheme}

Based upon the above conjecture I undertake explaining numerous
 charmonium-like states, which have unexpected and puzzling nature,
 in the $\widetilde{U}(12)_{SF} \times O(3,1)_{L}$-classification
 scheme. Here I will restrict myself to discussing the neutral
 charmonium-like $XY$ states which have large hadronic
 transition rates to lower conventional charmonia.
 The corresponding observed states, together with measured
 properties, are collected in Table \ref{tab:a}.


\begin{table}
\renewcommand{\arraystretch}{1.8}
\begin{tabular}{lccccc}
\hline
    \tablehead{1}{c}{b}{State}
  & \tablehead{1}{c}{b}{Mass (MeV)}
  & \tablehead{1}{c}{b}{Width (MeV)}
  & \tablehead{1}{c}{b}{$\bm{J^{PC}}$}
  & \tablehead{1}{c}{b}{Decay Modes}
  & \tablehead{1}{c}{b}{Experiments} \\
\hline
$Y(4008)$ & $4008^{+82}_{-49}$ & $226^{+97}_{-80}$ & $1^{--}$ & $\pi^{+}\pi^{-}J/\psi$
 & Belle \cite{Y(4008):Belle} \\
$Y(4260)$ & $4263^{+8}_{-9}$ & $95\pm 14$ & $1^{--}$ & $\pi^{+}\pi^{-}J/\psi$
 & Babar, CLEO, Belle \cite{PDG2009} \\
$Y(4360)$ & $4361\pm 13$ & $74\pm 18$ & $1^{--}$ & $\pi^{+}\pi^{-}\psi(2S)$
 & Babar, Belle \cite{Y(4360)Y(4660):Belle} \\
$Y(4660)$ & $4664\pm 12$ & $48\pm 15$ & $1^{--}$ & $\pi^{+}\pi^{-}\psi(2S)$
 & Belle \cite{Y(4360)Y(4660):Belle} \\
$X(3872)$ & $3871.46\pm 0.19$ & $<2.3$ & $1^{++}$
 & $\pi^{+}\pi^{-}J/\psi$, $\gamma J/\psi$, $D\bar{D}^{*}$
 & Belle, CDF, D0, Babar \cite{PDG2009,X(3872)} \\
$X(3915)$ & $3914\pm 4$ & $28^{+12}_{-14}$ & $0/2^{++}$ & $\omega J/\psi$
 & Belle \cite{X(3915):Belle} \\
$Y(3940)$ & $3916\pm 6$ & $40^{+18}_{-13}$ & $?^{?+}$ & $\omega J/\psi$
 & Belle, Babar \cite{PDG2009} \\
$Y(4140)$ & $4143.0\pm 3.1$ & $11.7^{+9.1}_{-6.2}$ & $?^{?+}$ & $\phi J/\psi$
 & CDF \cite{Y(4140):CDF} \\
$X(4350)$ & $4350.6^{+4.7}_{-5.1}$ & $13.3^{+18.4}_{-10.0}$ & $0/2^{++}$
 & $\phi J/\psi$ & Belle \cite{X(4350):Belle} \\
\hline
\end{tabular}
\caption{Measured properties of the neutral charmonium-like $XY$ states
 with large hadronic transition rates to lower conventional charmonia.
 The statistical and systematic errors of the respective masses and widths
 were added in quadrature.}
\label{tab:a}
\end{table}

To assign these nine charmonium-like states to $J^{PC}$ multiplets
 in this scheme, I make predictions on the mass spectrum of $1S$, $1P$,
 and $2S$ states for the $c\bar{c}$ system, assuming a phenomenological
 mass formula in which spin-dependent interactions are taken into account
 as
\begin{equation}
	M(n^{2S+1}L_{J}) = \overline{M}(nL; S^{(\rho)},S^{(\rho)}_{3})
	+ c_{LS}(nL)\langle \mathbf{L}\cdot \mathbf{S}\rangle
	+ c_{T}(nL)\langle S_{T}\rangle
	+ c_{SS}(nL)\langle \mathbf{S}_{Q}\cdot \mathbf{S}_{\bar{Q}}\rangle,
\end{equation}
where $\overline{M}(nL;S^{(\rho)},S^{(\rho)}_{3})$ are
 the spin-averaged masses of $1S$, $1P$,
 $2S$ states for the $(S^{(\rho)},S^{(\rho)}_{3})=(1,1)$, $(1,-1)$,
 $(1,0)$, $(0,0)$ sectors and $c_{LS}(nL)$, $c_{T}(nL)$, $c_{SS}(nL)$
 are the constant parameters which represent the contributions of
 spin-orbit, tensor, and spin-spin interactions, respectively.
 I take the excitation energies of $1P$ and $2S$ states to be equal
 for all the $(1,1)$, $(1,-1)$, $(1,0)$, $(0,0)$ sectors, so there are
 six independent parameters concerning the spin-averaged masses,
 $\overline{M}(1S)$, $\overline{M}(1P)$, $\overline{M}(2S)$ for the
 $(1,1)$ sector and $\overline{M}(1S)$'s for each of the $(1,-1)$,
 $(1,0)$, $(0,0)$ sectors. I also assume the following mass relations
 for the scalar and axial-vector states with $(1,0)$ and $(0,0)$: 
\begin{equation}
\begin{split}
	M(0^{++};S^{(\rho)}=1) &< M(0^{+-};S^{(\rho)}=0), \\
	M(1^{+-};S^{(\rho)}=1) &< M(1^{++};S^{(\rho)}=0),
\end{split}
\end{equation}
and the respective mass differences between the $S^{(\rho)}=0$ and
 $1$ states are identical.
The $c_{SS}(1S)$, $c_{SS}(2S)$, $c_{LS}(1P)$, $c_{T}(1P)$, and $c_{SS}(1P)$
 are taken to be common values for all the $(1,1)$, $(1,-1)$, $(1,0)$,
 $(0,0)$ sectors, except for the sign of $c_{SS}$ for the $(1,0)$ and
 $(0,0)$ sectors, and thus there are five independent parameters,
 $c_{SS}(1S)$, $c_{SS}(2S)$, $c_{LS}(1P)$, $c_{T}(1P)$, and $c_{SS}(1P)$.
 
To fix values of the model parameters I use measured masses
 of the $\eta_{c}(1S)$, $J/\psi(1S)$, $h_{c}(1P)$, $\chi_{c0,1,2}(1P)$,
 $\eta_{c}(2S)$, $\psi(2S)$ \cite{PDG2009},\footnote{The $\psi(2S)$ is 
 	assumed to be a pure $2^{3}S_{1}$ state, neglecting a small admixture
	of the $1^{3}D_{1}$ state.
 	}
 $X(3872)$ \cite{X(3872)},
 $X(3915)$ \cite{X(3915):Belle}, $X(4350)$ \cite{X(4350):Belle},
 assuming that the last two states have $J^{PC}=0^{++}$, and 
 extracted values are as follows:
%
%
\begin{equation}
\begin{split}
	&\overline{M}(1S;1,1) =3067.8 \ \text{MeV}, \ \
	 \overline{M}(1P;1,1) =3525.3 \ \text{MeV}, \ \
	  \overline{M}(2S;1,1) =3673.8 \ \text{MeV},\\
	&\overline{M}(1S;1,-1) =4003.7 \ \text{MeV}, \ \
	 \overline{M}(1S;1,0) =3826.6 \ \text{MeV}, \ \
	  \overline{M}(1S;0,0) =3901.5 \ \text{MeV}, 
\end{split}
\end{equation}
and
\begin{equation}
\begin{split}
	&c_{SS}(1S)=116.6 \ \text{MeV}, \ \ c_{SS}(2S)=49.0 \ \text{MeV}, \\
	&c_{LS}(1P)=35.0 \ \text{MeV}, \ \
	 c_{T}(1P)=40.6 \ \text{MeV}, \ \ c_{SS}(1P)\approx 0 \ \text{MeV}. 
\end{split}
\end{equation}
Using these values the charmonium spectrum of $1S$, $1P$, $2S$
 states are calculated, and then comparing the calculated masses of the
 predicted $J^{PC}$ states with the measured masses and $J^{PC}$ of 
 the relevant $XY$ states, I assign those $XY$ states to appropriate
 places. The results are given in Table \ref{tab:b}, where the
 ``$Y(4280)$'' is a hypothetical state mentioned by the CDF collaboration
 as follows \cite{Y(4140):CDF}: \textit{There is a small cluster of
 events approximately one pion mass higher than the first
 structure. However, the statistical significance of this
 cluster is less than $3\sigma$}. In Table \ref{tab:c} the suggested
 assignments for the relevant $XY$ states are summarized
 in comparison with experiment. From this table it is seen that
 the theoretical masses for the relevant $XY$ states are
 in excellent agreement with experiment.


\begin{table}
\renewcommand{\arraystretch}{1.8}
\begin{tabular}{ccccccccc}
\hline
  & \tablehead{1}{c}{b}{$P$}
  & \tablehead{1}{c}{b}{$V$}
  & \tablehead{1}{c}{b}{$P^{(\chi)}$}
  & \tablehead{1}{c}{b}{$V^{(\chi)}$}
  & \tablehead{1}{c}{b}{$S_{A}^{(\chi)}$}
  & \tablehead{1}{c}{b}{$B^{(\chi)}$}
  & \tablehead{1}{c}{b}{$S_{B}^{(\chi)}$}
  & \tablehead{1}{c}{b}{$A^{(\chi)}$} \\
\hline
$S^{(\rho)}$ & $1$ & $1$ & $1$ & $1$ & $1$ & $1$ & $0$ & $0$ \\
$S^{(\rho)}_{3}$ & $1$ & $1$ & $-1$ &$-1$ & $0$ & $0$ & $0$ & $0$ \\
\hline
  & $1^{1}S_{0}$ & $1^{3}S_{1}$ & $1^{1}S_{0}^{(\chi)}$ & $1^{3}S_{1}^{(\chi)}$
  & $1^{1}S_{0}^{(\chi A)}$ & $1^{3}S_{1}^{(\chi B)}$ & $1^{1}S_{0}^{(\chi B)}$
  & $1^{3}S_{1}^{(\chi A)}$ \\
$n=1$ & $0^{-+}$ & $1^{--}$ & $0^{-+}$ & $1^{--}$ & $0^{++}$ & $1^{+-}$
  & $0^{+-}$\tablenote{$J^{PC}$-exotic states.} & $1^{++}$ \\
\cline{2-9}
$L=0$ & $\underline{2980}$ & $\underline{3097}$ & $3916$ & $4033$ & $\underline{3914}$
  & $3797$ & $3989$ & $\underline{3872}$ \\
  & $\eta_{c}(1S)$ & $J/\psi(1S)$ & $Y(3940)$ & $Y(4008)$ & $X(3915)$ &
  &  & $X(3872)$ \\  
\hline
  &  & $1^{3}P_{0}$ &  & $1^{3}P_{0}^{(\chi)}$
  &  & $1^{3}P_{0}^{(\chi B)}$ &  & $1^{3}P_{0}^{(\chi A)}$ \\
  &  & $0^{++}$ &  & $0^{++}$ &  & $0^{-+}$
  &  & $0^{--*}$ \\
\cline{2-9}
  &  & $\underline{3415}$ &  & $\underline{4351}$ &  & $4144$ &  & $4248$ \\
  &  & $\chi_{c0}(1P)$ &  & $X(4350)$ &  & $Y(4140)$ &  & \\  
\cline{2-9}
\rule[14pt]{0pt}{0pt}
 & $1^{1}P_{1}$ & $1^{3}P_{1}$ & $1^{1}P_{1}^{(\chi)}$ & $1^{3}P_{1}^{(\chi)}$
  & $1^{1}P_{1}^{(\chi A)}$ & $1^{3}P_{1}^{(\chi B)}$ & $1^{1}P_{1}^{(\chi B)}$
  & $1^{3}P_{1}^{(\chi A)}$ \\
$n=1$ & $1^{+-}$ & $1^{++}$ & $1^{+-}$ & $1^{++}$ & $1^{--}$
  & $1^{-+*}$ & $1^{-+*}$ & $1^{--}$ \\
\cline{2-9}
$L=1$ & $\underline{3526}$ & $\underline{3511}$ & $4461$ & $4447$ & $4255$ & $4240$
  & $4359$ & $4344$ \\
  & $h_{c}(1P)$ & $\chi_{c1}(1P)$ &  &  & $Y(4260)$ &  &  & $Y(4360)$ \\  
\cline{2-9}
\rule[14pt]{0pt}{0pt}
  &  & $1^{3}P_{2}$ &  & $1^{3}P_{2}^{(\chi)}$
  &  & $1^{3}P_{2}^{(\chi B)}$ &  & $1^{3}P_{2}^{(\chi A)}$ \\
  &  & $2^{++}$ &  & $2^{++}$ &  & $2^{-+}$
  &  & $2^{--}$ \\
\cline{2-9}
  &  & $\underline{3556}$ &  & $4492$ &  & $4286$ &  & $4390$ \\
  &  & $\chi_{c2}(1P)$ &  &  &  & ``$Y(4280)$'' &  & \\  
\hline
  & $2^{1}S_{0}$ & $2^{3}S_{1}$ & $2^{1}S_{0}^{(\chi)}$ & $2^{3}S_{1}^{(\chi)}$
  & $2^{1}S_{0}^{(\chi A)}$ & $2^{3}S_{1}^{(\chi B)}$ & $2^{1}S_{0}^{(\chi B)}$
  & $2^{3}S_{1}^{(\chi A)}$ \\
$n=2$ & $0^{-+}$ & $1^{--}$ & $0^{-+}$ & $1^{--}$ & $0^{++}$ & $1^{+-}$
  & $0^{+-*}$ & $1^{++}$ \\
\cline{2-9}
$L=0$ & $\underline{3637}$ & $\underline{3686}$ & $4573$ & $4622$ & $4469$
  & $4420$ & $4544$ & $4495$ \\
  & $\eta_{c}(2S)$ & $\psi(2S)$ &  & $Y(4660)$ &  &  &  & \\  
\hline
\end{tabular}
\caption{Theoretical mass spectrum (in MeV) of conventional and chiralonic
 charmonia. The fitted mass values are underlined. The $n$ and $L$ are
 radial and orbital angular-momentum quantum numbers.}
\label{tab:b}
\end{table}


\begin{table}
\renewcommand{\arraystretch}{2.0}
\begin{tabular}{ccccccc}
\hline
  & \tablehead{2}{c}{b}{Experiment} &  & \tablehead{3}{c}{b}{Theory} \\
 \cline{2-3} \cline{5-7}
    \tablehead{1}{c}{b}{State}
  & \tablehead{1}{c}{b}{$\bm{J^{PC}}$}
  & \tablehead{1}{c}{b}{Mass (MeV)}
  & \tablehead{1}{c}{b}{}
  & \tablehead{1}{c}{b}{$\bm{J^{PC}}$}
  & \tablehead{1}{c}{b}{ Mass (MeV)}
  & \tablehead{1}{c}{b}{Assignment} \\
\hline
$Y(4008)$ & $1^{--}$ & $4008^{+82}_{-49}$
  &  & $1^{--}$ & $4033$ & $1^{3}S_{1}^{(\chi)}$ \\
$Y(4260)$ & $1^{--}$ & $4263^{+8}_{-9}$
  &  & $1^{--}$ & $4255$ & $1^{1}P_{1}^{(\chi A)}$ \\
$Y(4360)$ & $1^{--}$ & $4361\pm 13$
  &  & $1^{--}$ & $4344$ & $1^{3}P_{1}^{(\chi A)}$ \\
$Y(4660)$ & $1^{--}$ & $4664\pm 12$
 &  & $1^{--}$ & $4622$ & $2^{3}S_{1}^{(\chi)}$ \\
$X(3872)$ & $1^{++}$ & $3871.46\pm 0.19$
 &  & $1^{++}$ & $3872$ (fit) & $1^{3}S_{1}^{(\chi A)}$ \\
$X(3915)$ & $0/2^{++}$ & $3914\pm 4$
 &  & $0^{++}$ & $3914$ (fit) & $1^{1}S_{0}^{(\chi A)}$ \\
$Y(3940)$ & $?^{?+}$ & $3916\pm 6$
 &  & $0^{-+}$ & $3916$ & $1^{1}S_{0}^{(\chi)}$ \\
$Y(4140)$ & $?^{?+}$ & $4143.0\pm 3.1$
 &  & $0^{-+}$ & $4144$ & $1^{3}P_{0}^{(\chi B)}$ \\
$X(4350)$ & $0/2^{++}$ & $4350.6^{+4.7}_{-5.1}$
 &  & $0^{++}$ & $4351$ (fit) & $1^{3}P_{0}^{(\chi)}$ \\
``$Y(4280)$'' & $?^{?+}$ & $\approx 4280$
 &  & $2^{-+}$ & $4286$ & $1^{3}P_{2}^{(\chi B)}$ \\
\hline
\end{tabular}
\caption{Suggested assignments for the neutral charmonium-like
 $XY$ states in the $\widetilde{U}(12)_{SF} \times O(3,1)_{L}$-classification
 scheme in comparison with experiment.}
\label{tab:c}
\end{table}

As is seen from Table \ref{tab:b}, there exist some $J^{PC}$-exotic states,
 $0^{+-}(h_{c0})$ in the $1S$ and $2S$ levels and $0^{--}(\psi_{0})$,
 $1^{-+}(\eta_{c1})$
 in the $1P$ level,\footnote{
 	In the light $(u,d)$-quark sector there are presently experimental
	observations of the two exotic mesons with $I^{G}(J^{PC})=1^{-}(1^{-+})$,
	$\pi _{1}(1400)$ and $\pi _{1}(1600)$ \cite{PDG2009}, which can be
	assigned to the $1^{1}P_{1}^{(\chi B)}$ and $1^{3}P_{1}^{(\chi B)}$
	states.
	}
 which is a remarkable feature of the $\widetilde{U}(12)_{SF} \times
 O(3,1)_{L}$-classification scheme. Of particular interest is
 that a recent lattice QCD calculation of the charmonium spectrum showed the
 existence of exotic $1^{-+}(\eta_{c1})$ and $0^{+-}(h_{c0})$ states
 with masses of
 $4300(50)$ MeV and $4465(65)$ MeV, respectively \cite{Lattice:DR}.
 These masses might be
 compared with the corresponding predicted ones of $4240$ (or $4359$) MeV
 and $4544$ MeV for the $1^{-+}(1^{3}P_{1}^{(\chi B)} \ \textrm{or} \ 
 1^{1}P_{1}^{(\chi B)})$ and $0^{+-}(2^{1}S_{0}^{(\chi B)})$ states.
 A subsequent lattice QCD analysis on the radiative decay of the exotic
 $\eta_{c1}$ state at $4300(50)$ MeV, $\eta_{c1} \to J/\psi \gamma$,
 suggests that a $q\bar{q}$ pair in the $\eta_{c1}$ is in a spin
 triplet \cite{Lattice:DET} and thus it might be favorable to being assigned
 to $1^{3}P_{1}^{(\chi B)}$. Then the three states $Y(4140)$,
 ``$\eta_{c1}(\sim 4300)$'', ``$Y(4280)$'' would form the
 $1^{3}P_{J}^{(\chi B)}(0^{-+},1^{-+},2^{-+})$ multiplet.
 
In the present study another interesting prediction is obtained that there exists
 a spin-singlet partner, $\eta_{c}(2^{1}S_{0}^{(\chi)})$, of
 the $Y(4660)$, assigned above to $\psi(2^{3}S_{1}^{(\chi)})$, where the mass
 splitting between them is the same as that of the conventional charmonia,
 $\psi(2S)$ and $\eta_{c}(2S)$, being $\approx 49$ MeV.\footnote{
 The same prediction was given in a completely different approach
 of the hadronic molecule model with heavy-quark spin symmetry \cite{GHM2009}.}


\section{Final Remarks}

I have investigated the mass spectrum of
 $1S$, $1P$, and $2S$ states for the $c\bar{c}$ system in the
 $\widetilde{U}(12)_{SF} \times O(3,1)_{L}$-classification scheme and
 have seen that the observed spectrum of the neutral charmonium-like
 $XY$ states is described well. The resultant assignments
 suggest that the $J^{PC}$ quantum numbers of the $X(3915)$ and $X(4350)$
 are $0^{++}$, while both the $Y(3940)$ and $Y(4140)$ have $0^{-+}$, and
 there exists a charmonium-like state with a mass of $\approx4280$ MeV
 and $J^{PC}=2^{-+}$ which seems to correspond to the hypothetical state
 ``$Y(4280)$''. It is also expected that there exist some $J^{PC}$-exotic states,
 such as $1^{-+}(\eta_{c1})$ and $0^{+-}(h_{c0})$ states around $4.3$ GeV
 and $4.5$ GeV, respectively. 
 These results will be tested in the coming experiment.
 
In a future work, to put the present interpretation for the neutral
 charmonium-like $XY$ states on a firm basis, it is necessary to explain
 the decay properties of these $XY$ states that they do not decay dominantly
 into $D^{(*)}\bar{D}^{(*)}$, but have large hadronic transition rates to
 lower conventional charmonia.



\bibliographystyle{aipproc}   

\end{document}